\documentclass{article}

\usepackage{amssymb, amsmath, amsthm, verbatim, graphicx}

\begin{document}

\title{Constructing Lower-Bounds for CTL Escape Rates in Early SIV Infection}
\author{Sivan Leviyang\\ 
Georgetown University\\  
Department of Mathematics and Statistics}
\maketitle
  
\abstract{Intrahost simian immunodeficiency virus (SIV) evolution is marked by repeated viral escape from cytotoxic T-lymphocyte (CTLs) response.  Typically, the first such CTL escape occurs in a matter of days, starting around the time of peak viral load.   Many authors have developed methods to quantify the strength of CTL response by measuring the rate at which CTL escape occurs, but such methods usually depend on sampling at two or more timepoints, while many datasets capture the dynamics of the first CTL escape at only a single timepoint.  Here, we develop inference methods for CTL escape rates applicable to single timepoint datasets.   Through a model of early infection dynamics, we construct confidence intervals for escape rates, but since early infection dynamics are not completely understood, we also develop a one-sided confidence interval serving as a lower bound for escape rates over a collection of early infection models.  We apply our methods to two SIV datasets, using our lower bounds and existing methods to show that escape rates are relatively high during the initial days of the first CTL escape and then drop to lower levels as the escape proceeds.   We also compare escape in the lymph nodes and the rectal mucosa, showing that escape in the lymph nodes is initially faster, but as the first escape proceeds, the rate of escape in the lymph nodes drops below the rate seen in the rectal mucosa.}

\renewcommand{\min}{{\text{min}}}
\renewcommand{\max}{{\text{max}}}
\newcommand{\LB}{\text{LB}}
\newcommand{\bark}{\bar{k}}
\newcommand{\wm}{\frac{w(t_A)}{m(t_A)}}
\newcommand{\mw}{\frac{m(t_A)}{w(t_A)}}
\newcommand{\pwm}{\left( \wm \right)}
\newcommand{\pmw}{\left( \mw \right)}
\renewcommand{\P}{\mathcal{P}}
\newcommand{\p}{p}
\newcommand{\kp}{(\bark)}
\newcommand{\model}{{\text{model}}}
\newcommand{\low}{{\text{lower}}}
\newcommand{\actual}{{\text{actual}}}
\newcommand{\twm}{\tilde{\wm}}
\newcommand{\W}{\mathbb{W}}

\renewcommand{\beta}{\xi}
\renewcommand{\alpha}{{\xi_E}}
\newcommand{\CI}{{\text{CI}}}
\newcommand{\fit}{\text{fit}}
\renewcommand{\c}{c_{\fit}}

\section{Introduction}

	During HIV and SIV infections, the viral population repeatedly escapes from selective pressure exerted by cytotoxic T-lymphocytes (CTLs), a type of immune system cell.  Each CTL targets a specific peptide, referred to as an epitope, associated with a locus on the viral genome.  Mutation at the locus may change the epitope, making it partially or completely unrecognizable by existing CTLs.  Cells infected by a viral genome possessing such mutations are at a selective advantage, leading to a selective sweep referred to as a CTL escape.  See \cite{Goulder_Nature_Reviews_2004} for a review of CTL escape in both HIV and SIV infection.
	
	In this work, we consider the first CTL escape to occur during an infection.   In SIV and HIV infection, CTL response initiates roughly at 14 and 21 days after infection, respectively, just prior to peak viral load \cite{Borrow_1994_J_Virol, Cohen_2011_NJM, Goulder_Nature_Reviews_2004, McMichael_2010_Nat_Rev_Imm}.   In the week or two following the initiation of CTL response, CTL escape often occurs at a single targeted epitope \cite{Boutwell_2010_J_Infec_Dis, McMichael_2010_Nat_Rev_Imm, Goonetilleke_2009_JEM, Henn_2012_PLOS_Path, Allen_2000_Nature}.   T-cell tetramer studies suggest that this escape is driven by an especially focused CTL response in comparison to subsequent responses and escapes \cite{Turnbull_2009_J_Immunology, Yasutomi_1993_J_Virol, Veazey_2003_J_Med_Primat}.
	
	Many authors have attempted to quantify the strength of CTL response by measuring the rate at which CTL escape occurs.  A commonly used method (e.g. \cite{Goonetilleke_2009_JEM, Love_2008_J_Virol, Loh_2008_PLOS_Pathogens, Asquith_PNAS_2007, Ganusov_2011_J_Virology}), introduced in \cite{Fernandex_J_Virol_2005, Asquith_PLOS_Biology_2006}, fits escape mutation frequencies at two timepoints to a differential equation model.  The model fit is determined by a single parameter, known as the escape rate, which is used to quantify the strength of CTL response at a given epitope.  Since this approach requires frequency data at two timepoints, we call it the two-point method.  
	
	Using the two-point method to analyze the first CTL escape is difficult because rarely do both sampled timepoints capture the escape.  For example, the first two timepoints available in HIV studies of acute infection are typically in the range of day 30 and 50, e.g. \cite{Fisher_2010_PLOS_One, Goonetilleke_2009_JEM}.  Using the two-point method on such data estimates escape rates between day 30 and 50, while CTL response is likely strongest prior to day 30.  The situation is different for SIV studies.  Since the time of infection can be controlled, sampling timepoints can be chosen that straddle day 14, the approximate time of CTL response; for example sampling can occur at days 7 and 21.   But usually the CTL escape has not started at day 7, so the two-point method must be applied using data collected at day 21 and a later timepoint, leading to the same difficulties seen in HIV datasets.  Even when the first escape is caught twice, say at days 14 and 18, the escape rate prior to day 14 cannot be inferred using the two-point method. Further, when sampling times become too close, the effect of sampling variance leads to wide confidence intervals in the two-point method.
	
	Other authors  (e.g. \cite{Mandl_J_Virol_2007, Althaus_PLOS_Comp_Bio_2008, Petravic_2008_J_Virol, Monteiro_2000_JTB}) have developed methods based on the standard model of viral dynamics \cite{Perelson_Nature_Reviews_2002, Nowak_and_May_Book}.  These methods depend on models with many parameters, in contrast the two-point method has no parameters besides the estimated escape rate.  Further, fitting of the standard model and its variants requires multiple timepoints, so that the time period to which such escape rate estimates apply is often unclear.
	
	The uniqueness of the first CTL escape, in terms of its temporal association with peak viral load and the focused CTL response that drives it, makes its quantification biologically valuable.  But further, the first escape is free of complexitites that make later CTL escapes difficult to model.   Soon after peak viral load, in both HIV and SIV, multiple CTL escapes occur, often overlapping in time (e.g. \cite{Bimber_2010_Nature, Henn_2012_PLOS_Path, Goonetilleke_2009_JEM}, see \cite{Boutwell_2010_J_Infec_Dis} for a review).   The interaction of viral variants involved in such sweeps, both through inter-variant competition for target cells and possible recombination events, makes modeling and inference complex \cite{Leviyang_2013_Genetics, Neher_2010_PLOS_Comp_Bio, Kessinger_2013_Frontiers_T_Cell, Batorsky_2011_PNAS}.   From this view, the first CTL escape may be an ideal setting in which to infer escape rates.

	The rate of CTL escape can be defined in different ways.  For example, some authors measure the timespan from initiation of CTL response to the time when mutant frequencies reach a prescribed level \cite{Liu_2006_J_Virology, Palmer_2013_Proc_Roy_Soc_B}.  In the two-point method, using the underlying model, the escape rate is the sum of the average CTL kill rate and the fitness cost of mutation \cite{Asquith_PLOS_Biology_2006, Fernandex_J_Virol_2005}.  We take this as our definition of the escape rate.
	
	 In this work, we develop inference methods for estimating the rate of the first CTL escape using frequency data from a single timepoint.  We apply these methods to SIV datasets, a setting in which inference is slightly easier because infection time is typically known.  We have in mind frequency data collected somewhere between days 14 and 28, times that capture the first CTL escape when the mutation frequency is substantial, but before escape at other epitopes has developed.
	  
	  The price we pay for using a single timepoint is the addition of three parameters: $\mu$, the rate at which mutations accumulate in the epitope; $t_A$, the time at which CTL response initiates;  and $P_A$, the number of cells infected at time $t_A$.   We build two confidence intervals (CIs) for the escape rate.   First, we build a model-based CI, making assumptions about the infected cell growth rate  and the time dependence of the CTL response.   The aim of the model-based CI is to estimate the escape rate using our best guess of early SIV and CTL dynamics.    Second, we build a lower-bound CI.   This is a one-sided CI of the form $[e_\low,\infty)$ that serves as a lower bound for the escape rate.   In this setting we do not choose a model for SIV and CTL dynamics, rather we construct a CI that applies across a range of possible models.  
	  
	The early dynamics of SIV infection are not completely understood, for instance the number of infected cells during the eclipse phase is unknown.   To deal with this uncertainty, we specify a range of possible models, and the lower-bound CI contains the CI of all models in this range.   In many cases, as we show below, the lower-bound CI is uninformative; for example when it contains all escape rates greater than zero.  But often the lower-bound CI is informative, allowing us to make inferences that are relatively free of model dependence. 

	We apply our methods to two SIV datasets: Bimber et. al. \cite{Bimber_2009_J_Virol} and Vanderford et. al. \cite{Vanderford_2011_PLOS_Pathogens}.	  For Bimber et. al.,
 we use our methods, applied to day 21 data, and the two-point method, applied to day 21 and 28 data, to show the escape rate prior to day 21 is greater than the rate during days 21 to 28.  Vanderford et. al. includes mutation frequency data from several compartments, in particular lymph nodes and the rectal mucosa.   As we did for Bimber et. al, we consider two timepoints, in this case days 14 and 28, applying our method to day 14 data and the two-point method to day 14 and 28 data.  We are able to compare escape rates across different times and compartments, showing that escape in the lymph nodes prior to day 14 is higher than during days 14 to 28, while the escape rate in the rectal mucosa is less than pre-day 14 rates but greater than post-day 14 rates in the lymph nodes.     
	
	Mathematical details relating to construction of the model-based and lower-bound CIs are found in the Methods sections.  Readers interested only in biological implications can read the Results and Discussion sections. 
	
\section{Results}

	Most existing analyses of CTL escape infer a single rate for the entire escape, but by combining our methods with the existing two-point method, rates can be estimated for two separate time periods of the escape.   We apply our CI construction to an early sampled timepoint, day 21 for Bimber et. al and day 14 for Vanderford et. al., to infer the escape rate during the early portion of the first CTL escape;  then we apply the two-point method using that same timepoint and a later timepoint, days 21 and 28 for Bimber et. al. and days 14 and 28 for Vanderford et. al, to infer the escape rate during the latter portion of the CTL escape.   Importantly, although our methods use a single data timepoint, the escape rate estimate applies to the time interval starting at $t_A$, the time at which CTL response begins, and ending at the data timepoint.  
	
	To build CIs, we must specify the three parameters $\mu, t_A, P_A$.   We set $\mu = 3 * 10^{-4}$, roughly $10$ times the standard, per base pair, per replication cycle, SIV mutation rate \cite{Mansky_1996_Aids_Res_Hum_Retro}.   Our $\mu$ represents the rate at which the epitope mutates, not the mutation rate of a single base pair.   TAT-SL8 and NEF-RM9 are the epitopes considered below and both datasets reveal roughly $10$ different epitope mutations during escapes.  Mandl et. al. \cite{Mandl_J_Virol_2007} used a similar mutation rate to analyze TAT-SL8 escape.   $t_A$, the initial time of CTL response, and $P_A$, the number of infected cells at time $t_A$, differ between the datasets and are specified below. 	
	
	In some of the following figures, the CIs include negative escape rates.  Briefly, negative escape rates result from accounting for sampling variance.  For example, the mutation frequencies provided by the data might be $.4$ and $.45$ at two subsequent timepoints, but sampling variance allows the true mutation frequencies from which the data was sampled to be, say, $.43$ and $.42$ at the subsequent timepoints.  If the drop in mutation frequencies over time has significant likelihood given the data, the CI would include a negative escape rate.

\subsection{Bimber et. al. dataset}

	The data from Bimber et. al. involves four Mauritian cynomolgus macaques (MCMs) and four Rhesus macaques (RMs).  (The full dataset included eight RMs, we considered the four unvaccinated RMs.)   We refer the reader to the article and references therein for full details \cite{Bimber_2009_J_Virol}.  Briefly, all animals were intrarectally infected with SIVmac239.    The first CTL escape in the MCMs was at the epitope NEF-RM9, while CTL escape first occurred at TAT-SL8 in the RMs.    Pyrosequencing of the epitopes was performed at various timepoints.   At day 14 after infection, for both MCMs and RMs, the sampled sequences were roughly homogeneous; by day 21, MCMs and RMs had a significant frequency of escape mutants at NEF-RM9 and TAT-SL8, respectively.   We assumed that CTL response arose at day 14, i.e. $t_A = 14$, reflecting tetramer data showing CTL response arising at that time.  $P_A$, the number of infected cells at time $t_A$, has only a minor effect on the results.  We tried values ranging from $10^6$ to $10^{10}$ with little difference in outcome. Results shown were produced using $P_A = 10^8$. 
	
\begin{figure} [h]
\begin{center} 
\includegraphics[width=1.2\textwidth]{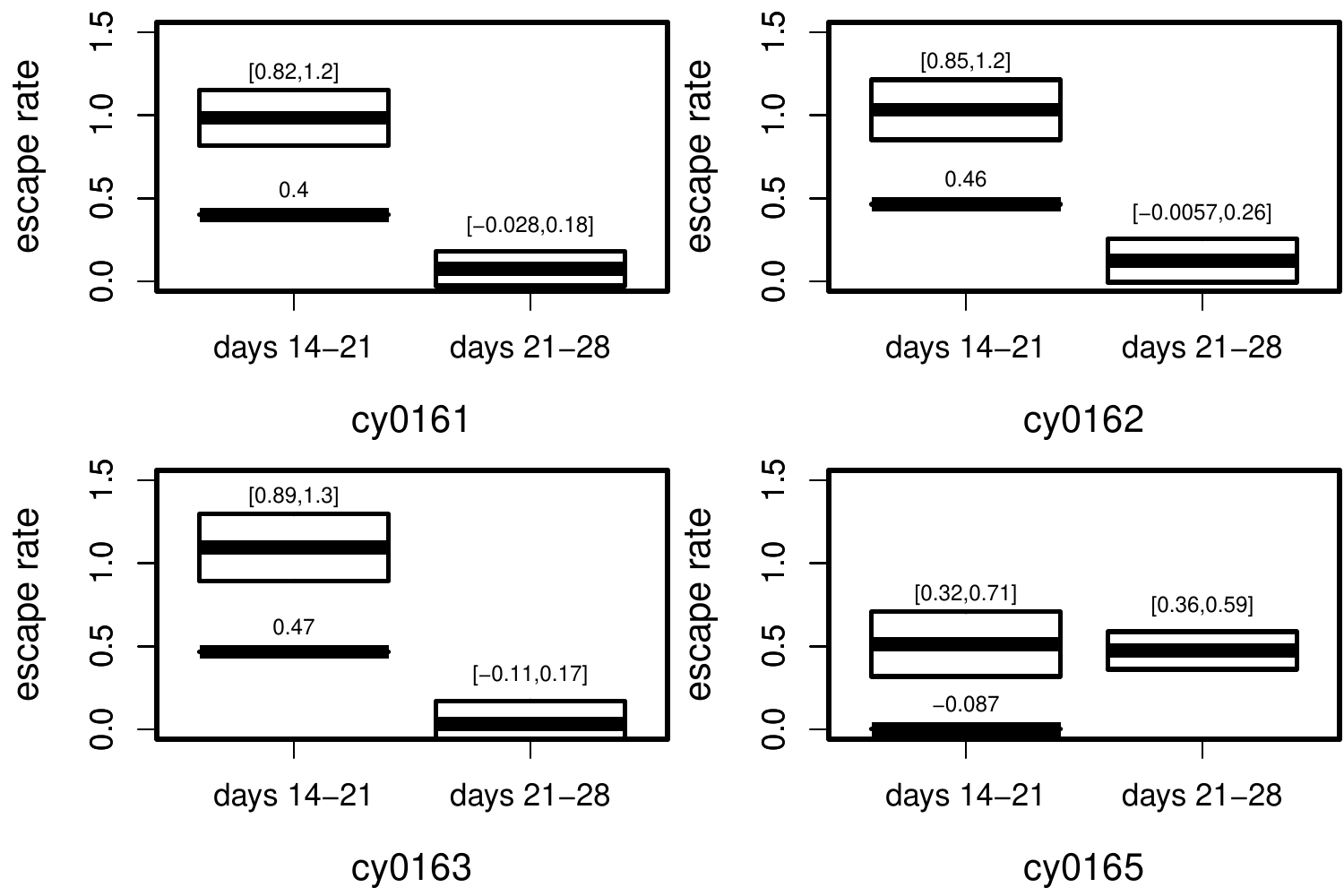}
\caption{Escape Rates CIs for MCMs in Bimber et. al.  Each subfigure represents a single animal.  Within each subfigure, the left tic gives escape rate CIs for days 14-21, with the model-based CI (the box) above the left endpoint of the lower-bound CI (the darkened line).  The right tic gives the escape rate CI for days 21-28 as computed by the two-point method.   All CIs are at $95\%$ significant and include sampling variance.} 
\label{F:MCM}
\end{center}
\end{figure}
	
	Figure \ref{F:MCM} shows results for the four MCMs.  In each subfigure, the left tic gives our model-based and lower-bound CIs while the right tic gives the two-point CI.   As the figure shows, in three of the four animals, the escape rate during days 14-21 appears to be higher;  not only are the model-based CIs greater than the two-point method CIs, but the lower-bound CIs are greater as well.  Interestingly, animal cy0165, which is the only animal with overlapping two-point and lower-bound CIs, had a weak CD8+ response from day 14 to 21 and a response that was increasing shortly after day 21 (see Figure 3 in Bimber et. al.).

\begin{figure} [h]
\begin{center} 
\includegraphics[width=1.2\textwidth]{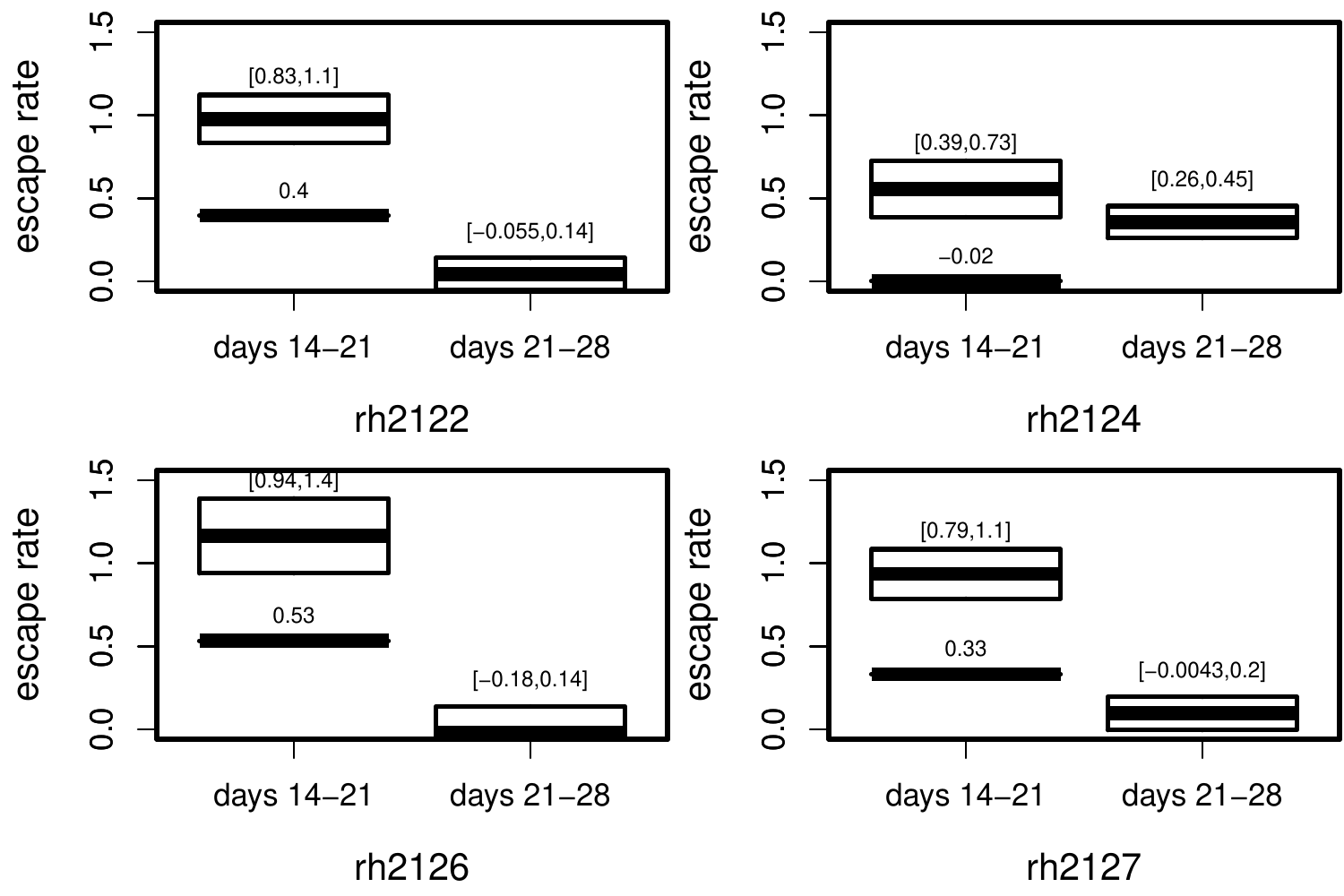}
\caption{Escape Rates CIs for RMs in Bimber et. al.  See Figure \ref{F:MCM} for details.} 
\label{F:RM}
\end{center}
\end{figure}
	
	Figure \ref{F:RM} shows analogous results for the four RMs.  Like the MCMs, for one RM the lower-bound CI is below the two-point CI, unfortunately no tetramer data is available for comparison.   
	
\subsection{Vanderford et. al. dataset}

	The Vanderford et. al. dataset includes fifteen Rhesus macaques (RMs) infected intravenously with SIVmac239.   As in Bimber et. al., the RMs experienced initial escape at TAT-SL8 and escape dynamics were tracked using pyrosequencing.   We refer the reader to the article and references therein for full details \cite{Vanderford_2011_PLOS_Pathogens}, here we only mention those aspects of the dataset relevant to our analysis.  Escape dynamics were tracked in four compartments : viral RNA in the plasma (PL) and genomic DNA from peripheral blood mononuclear cells (PBMC), lymph node biopsies (LN), and rectal mucosa biopsies (RB).  Sequencing was performed at days 14 and 28 as well as other later timepoints.   Using tetramer data, the frequency of CD8+ T-lymphocytes  specific for TAT-SL8 was estimated in the different compartments at days 7, 14 and 28.
		
	Vanderford et. al. show that lymph nodes and rectal mucosa are the primary source of TAT-SL8 escape mutants, with escape mutants often first arising in the lymph nodes.  Given this result, we focused on inferring rates of escape in the LN and RB compartments.  In order to consider only escape starting in lymph nodes, we restricted our attention to animals in which PL, PBMC, and RB epitope frequencies were above $90\%$ at day 14 and for which LN escape data was available at days 14 and 28 (not all animals were sampled at all timepoints).   Six RMs fulfilled these requirements.    
	
	   Tetramer data for the LN showed a weak CD8+ response at day 7 but a strong response by day 14.  Given this data, we assumed that significant CTL killing started at day 9, meaning that we set $t_A = 9$.  Choosing $t_A > 9$ would increase escape rate estimates, leaving our conclusions unchanged.   For RB, we assumed CTL killing started at day 14, also reflecting tetramer data.  
	   
	   For RB, we set $P_A$, the number of cells infected at time $t_A$, equal to  $10^8$.  For LN, since day 9 significantly precedes peak viral load, we hypothesized that in the absence of CTL response the LN would have $10^7$ infected cells at day 14 and then interpolated the number of infected cells at day 9 based on a constant growth rate; this led to $P_A$ of roughly $32000$.   Raising the $P_A$ values by factors of up to $100$ had little effect on our results.  
	  
\begin{figure} [h]
\begin{center} 
\includegraphics[width=1.2\textwidth]{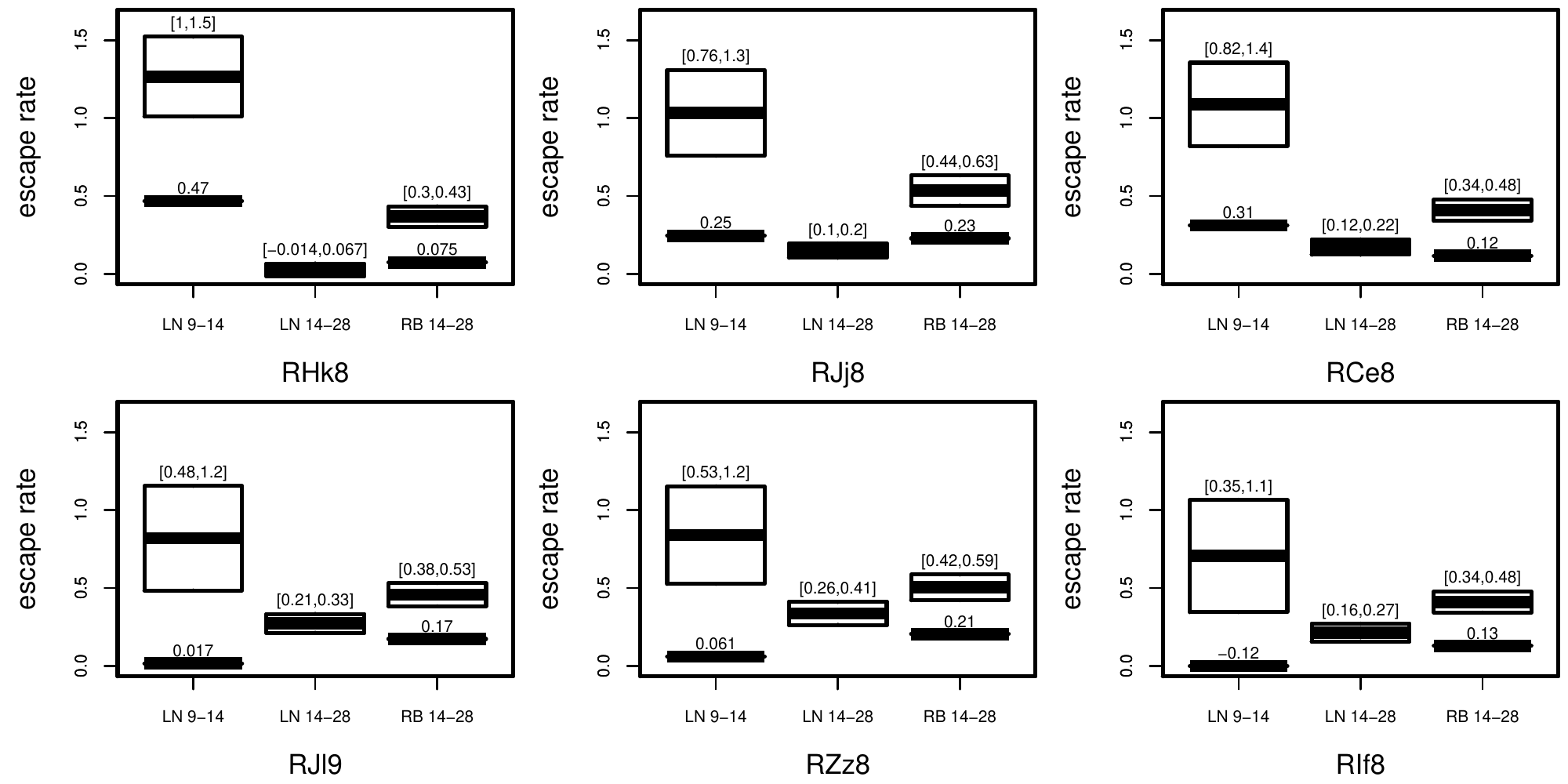}
\caption{Escape Rates CIs for Vanderford et. al.  Each subfigure represents a single animal.  Tics, from left to right, represent escape rates in the lymph node during days 9-14, lymph nodes during days 14-28, and rectal biopsies during days 14-28.   Within each subfigure, the left-most and right-most tics show the model-based CI (the box) above the left endpoint of the lower-bound CI (the darkened line).  All CIs are at $95\%$ significance.} 
\label{F:Vand_model}
\end{center}
\end{figure}

	Figure \ref{F:Vand_model} gives confidence intervals for the escape rates in the six RMs.  Each subfigure corresponds to a single animal. The first and second tics represent the escape rate in the LN during days 9-14, constructed using our methods, and days 14-28, constructed using the two-point method, respectively.  The third tic represents the escape rate in RB during days 14-28, constructed using our methods and day 28 data; application of the two-point method is not possible because escape in the RB has just started at day 14.
	
	Across all animals the model based CIs predict a significantly higher LN escape rate during days 9-14 than escape rates inferred for days 14-28 using the two-point method.  This pattern is supported by the lower-bound CIs for the RMs represented in the top row, but the lower-bound CIs in the bottom row are poor and provide little information.
	
	Similarly, across all animals, the model-based CIs predict higher RB escape rates during days 14-28 than those seen in the LN during the same period.   In this case, the lower-bound CIs for the RB escape rate are either slightly above or below the LN escape rates, so we can conclude that RB escape rates during days 14-28 are at least roughly the same as those experienced in the LN during that time.   Based on numerical experiments (see below), lower-bound CIs are conservative, meaning that RB escape rates are likely higher than LN escape rates during days 14-28.

\section{Methods}

	Our model and inference methods distinguish between two types of infected cells: wild type and mutants.  Wild types contain the epitope at which first CTL escape occurs, while mutants contain a nucleotide mutation at that epitope.   $w(t)$ and $m(t)$ represent the number of wild type and mutant infected cells, respectively, at time $t$.
	
	We construct our CIs under a model assuming wild type and mutants have equal fitness.  However, the CIs are applicable in the presence of fitness differences, as we show in subsection \ref{S:fitness}.

	To construct the lower-bound CI, we use the idea of a stochastic bound.  For us, technically, a random variable $Y$ is a stochastic upper bound of $X$ if both are defined on the same probability space $\Omega$ and for $\omega \in \Omega$, $X(\omega) \le Y(\omega)$; see, for example, \cite{Lindvall_Coupling_Book} for more details and examples.   A simple example, to orient the unfamiliar reader, is the case of $X$ as the result of a die roll, defined by a uniform probability of rolling $1$ through $6$, and the stochastic upper bound $Y$ taking the value, say, $4$ on die rolls $1,2,3$ and, say, $9$ on die rolls $4,5,6$.  In the development below, we strive to explain our stochastic bounds without excessive technicalities.
	
\subsection{Model}  \label{S:model}

	In this subsection, we specify the model on which our inference is based.   For the model-based CI, we choose a specific parametrization of this model, while for the lower-bound CI we specify and bound a range of parametrizations.
	
	The dynamics of $w(t)$ are parametrized by the growth rate $r(t)$, the attack time $t_A$, the CTL kill rate $k(t)$, and $P_A$, the number of infected cells at time $t_A$.   The attack time identifies the beginning of CTL response, which formally means that $k(t) = 0$ for $t < t_A$.  Wild type dynamics are then given by,
\begin{equation} \label{E:w_dynamics}
w(t) = \bigg\{
\begin{array}{cc}
\exp[\int_0^t ds r(s)] & \text{for } t \le t_A \\
P_A \exp[\int_{t_A}^t ds (r(s) - k(s))] & \text{for } t > t_A \\
\end{array}
\end{equation}
Note that we assume $w(0) = 1$, although this is not essential.  Further, we have the consistency condition for $r(t)$ and $P_A$,
\begin{equation}  \label{E:P_A_constraint}
P_A = \exp[\int_0^{t_A} ds r(s)].
\end{equation}

	The evolution of mutants is parametrized by a mutation rate $\mu$ and an offspring distribution $X(t;t_0)$. Mutants arise from wild types at rate $\mu$.   $X(t; t_0)$ is the number of infected cells at time $t$ that arise from a single infected cell that experiences an epitope mutation event at time $t_0$.  $X(t; t_0)$ is a random variable for fixed $t, t_0$.  Assuming no fitness differences between wild types and mutants gives the condition
\begin{align} \label{E:X_E}
E[X(t;t_0)] & = \exp[\int_{t_0}^t ds r(s)]
\\ \notag
	& =
\left(\frac{w(t)}{w(t_0)}\right) \exp[\int_{t_0}^t ds k(s)],
\end{align}
and notice if $t < t_A$ then $E[X(t; t_0)] = w(t)/w(t_0)$.   

We define $m(t)$ by
\begin{align} \label{E:m_dynamics}
m(t) & = \int_0^t P(\mu w(s) ds) X(t; s),
\end{align}
where $P(\mu w(s))$ is a Poisson process run at rate $\mu w(s)$, meaning that a mutation occurs during $[s, s+\Delta s]$ with probability $\mu w(s) \Delta s$.  When such a mutation occurs, the random variable $X(t; s)$ is sampled to determine the number of mutant infected cells at time $t$ that descend from this particular mutation event.  We make no assumption on the form of the offspring distribution $X(t;t_0)$, aside from the constraint on $E[X(t;t_0)]$ mentioned above.  However, in (\ref{E:m_dynamics}) we assume each mutation event samples $X(t;s)$ independently.

\subsection{Inference}

	For the sake of clarity, we first explain our inference methods assuming wild types and mutants are equally fit.   In subsection \ref{S:fitness} we consider the case of unequal fitness. 
	
	Let $t_F$ be the sampling time and $\hat{f}$ the mutant frequency sampled.  Define $f(t)$ as the fraction of infected cells of mutant type at time $t$, i.e. $f(t) = m(t)/(w(t) + m(t))$, $\hat{f}$ is formed by sampling a population with frequency $f(t_F)$.   More formally, $\hat{f}$ is a sample from the distribution $\tilde{f}$ defined by,
\begin{equation} \label{E:tilde_f}
\tilde{f} = \text{Binomial}(f(t_F), n),
\end{equation}
where $\text{Binomial}(a,b)$ is a binomial distribution with success probability $a$ and $b$ trials, and $n$ is the number of viral sequences sampled at $t_F$. 

	We assume that $f(t_F)$ is generated by our model under a specific parametrization, meaning a choice for $\mu, t_A, P_A$ and $r(t),k(t),X(t;t_0)$.  Let $\bark$ be the average of $k(t)$ over $[t_A, t_F]$, our goal is to construct confidence intervals for $\bark$ given the data $\hat{f}$.   Since we assume no fitness costs, $\bark$ is the escape rate. Given the high dimensional parameter space, our approach is to assume $\mu, t_A$ and $P_A$ are given and then construct a model-based and lower-bound CI for $\bark$.  To construct the model-based CI we make specific choices, described below, for $r(t)$, $k(t)$, and $X(t;t_0)$.   The lower-bound CI applies over a range of possible $r(t), k(t)$ and $X(t; t_0)$ and is not associated with a certain parameter choice.    
	
	  Our CI constructions involve two steps.
\begin{enumerate}
\item Based on the model of subsection \ref{S:model}, we determine the distribution of the mutant frequency at time $t_A$.
\item Given the mutant frequency distribution at time $t_A$ and the data at time $t_F$, we use a generalized two-point method to estimate the escape rate during $[t_A,t_F]$.
\end{enumerate}
Since the mutant frequency at time $t_A$ is small, of order $\mu$, the escape during $[t_A, t_F]$ is still partly driven by mutation.    Applying the two-point method to the frequencies at times $t_A$ and $t_F$ would ignore this effect, resulting in under-estimation of escape rates; so instead we use a generalized two-point method which accounts for mutation.

	We decompose $f(t_F)$ based on the two steps mentioned above.  From (\ref{E:w_dynamics}) and (\ref{E:m_dynamics}), simple algebra gives
\begin{equation}  \label{E:basic_inference_0}
f(t_F) = \frac{1}{1 + \frac{z}{\Gamma + z\beta}},
\end{equation}
where
\begin{equation}  \label{E:gamma_def}
\Gamma = \mw,
\end{equation}
\begin{equation} \label{E:z_def}
z = \exp[-\bark(t_F - t_A)],
\end{equation}
and
\begin{equation}
\beta = \int_{t_A}^{t_F} P(\mu w(s) ds) \frac{X(t_F;s)}{w(t_F)}.
\end{equation}
$\Gamma$ is essentially the frequency of mutants at time $t_A$, reflecting the first step of our CI construction.   If we set $\mu = 0$, then $\beta = 0$ and solving (\ref{E:basic_inference_0}) for $\bark$ reduces to the two-point method if $\Gamma$ is a fixed scalar.   However when $\Gamma$ is small, mutation and the dynamics of $k(t)$ during $[t_A, t_F]$ affect mutant frequency at time $t_F$ in a significant way. 

	For the purposes of inference, we replace $\beta$ in (\ref{E:basic_inference_0}) by its expected value, $\alpha$.  Taking expected values of a Poisson process, replaces $P(\mu w(s) ds)$ by $\mu w(s) ds$ and then (\ref{E:X_E}) and simple algebra give
\begin{equation} \label{E:alpha_def}
\alpha = \mu \int_{t_A}^{t_F} ds \exp[\int_s^{t_F} ds' k(s')],
\end{equation}
and the approximation
\begin{equation}  \label{E:basic_inference}
f(t_F) \approx \frac{1}{1 + \frac{z}{\Gamma + z \alpha}}.
\end{equation}
(\ref{E:basic_inference}) is the two-point method generalization mentioned in step two, above.  

	In SIV infection, after time $t_A$,  $m(t)$ is on the order of $1000$s and $\mu w(t)$ is on the order of $100$s.   In this regime, mutations occur quickly and averaging effects allow us to replace $\beta$ by its expected value $\alpha$ with small error.  Importantly, in (\ref{E:basic_inference}) only $\Gamma$ depends on $r(t)$ and then only when $t < t_A$; since $r(t)$ is not well understood, especially near and after peak viral load, basing inference only on $r(t)$ dynamics prior to $t_A$ is a significant, modeling advantage.

	To define our CIs, we let $\P(\mu, t_A, P_A, \bark)$ be the set of $r(t), k(t), X(t;t_0)$ that we assume possible for early SIV dynamics and CTL response given $t_A, P_A, \bark$.  $\P(\mu, t_A, P_A, \bark)$ is defined through a collection of constraints.  First, $r(t), k(t), X(t;t_0) \in \P(\mu, t_A, P_A, \bark)$ must satisfy the definitions and assumptions already mentioned:
\begin{itemize}
\item $k(t)$ has average $\bark$ over the interval $[t_A, t_F]$,
\item $r(t)$ satisfies (\ref{E:P_A_constraint}),
\item $E[X(t;t_0)] = w(t)/w(t_0)$ for $t_0, t \in [0, t_A]$.
\end{itemize}
We place two additional constraints on $r(t)$ and $k(t)$ motivated by our desire to construct a lower-bound CI.  The specific form of these constraints is given subsection \ref{S:lower_bound}, where their technical motivation is made clear.

	With $\P(\mu, t_A, P_A, \bark)$ defined, we can consider functions mapping $\bark \in \mathbb{R}$ to values in $\P(\mu, t_A, P_A, \bark)$, so that every $\bark$ is associated with a $r(t), k(t), X(t;t_0)$.   Letting $\p(\bark)$ be such a function;  $z$, $\alpha$ and the distribution of $\Gamma$ in (\ref{E:basic_inference}) can also be thought of as functions of $\bark$ which we write as $z_\p(\bark)$, $\alpha_\p(\bark)$ and $\Gamma_\p(\bark)$.  In turn, using $\p(\bark)$ in (\ref{E:basic_inference}) and plugging the result into (\ref{E:tilde_f}) leads to the definition
\begin{equation}  \label{E:tilde_f_CI}
\tilde{f}^\CI_\p(\bark) 
= \text{Binomial}(\frac{1}{1 + \frac{z_\p(\bark)}{\Gamma_\p(\bark) + z_\p(\bark)\alpha_\p(\bark)}},n).
\end{equation}
We can form a CI by determining the interval of $\bark$ for which $\hat{f}$ has a p-value of greater than $.05$ under $\tilde{f}_\p^\CI(\bark)$.  

	To orient the reader, we contrast the distributions $\tilde{f}$ and $\tilde{f}^\CI_\p(\bark)$.   $\tilde{f}$ is binomially distributed with success probability $f(t_F)$ and the distribution of $f(t_F)$ is parameterized by the given values of $\mu, t_A, P_A$ and specific choices for $r(t), k(t), X(t;t_0)$.  In contrast, $\tilde{f}^\CI_\p(\bark)$ is binomially distributed, but with success probability given by the approximation (\ref{E:basic_inference}) of $f(t_F)$; the distribution of (\ref{E:basic_inference}) is paramaterized by $\bark$, the given values of $\mu, t_A, P_A$ and through $\p(\bark)$.

	Let $[e_{\model,1}, e_{\model,2}]$ and $[e_\low,\infty)$  be the model-based and lower-bound CI, respectively. To form the model-based CI, we use precisely the approach of the previous paragraph, defining a mapping $\p_\model(\bark)$ which we decompose as,
\begin{equation}  \label{E:p_model}
\p_\model(\bark) = (r_\model^{\kp}(t), k_\model^{\kp}(t), X_\model^{\kp}(t)).
\end{equation}
The precise form of $p_\model(\bark)$ is given below, in subsection \ref{S:model_based}.  Then, defining $\tilde{f}_\model^\CI$ by setting $\p = \p_\model$ in (\ref{E:tilde_f_CI}),
\begin{align} 
e_{\model,1} & = \inf \{\bark : P(\tilde{f}_\model^\CI(\bark) \ge \hat{f}) \ge .025)\},
\\ \notag
e_{\model,2} & = \sup \{\bark : P(\tilde{f}_\model^\CI(\bark) \le \hat{f}) \ge .025)\}.
\end{align}

	To form the lower-bound CI, we do not build a function similar to $\p_\model(\bark)$.  Instead, we directly define $\alpha_\low(\bark)$ and $\Gamma_\low(\bark)$ which replace $\alpha$ and $\Gamma$ in (\ref{E:tilde_f_CI}), thereby defining $\tilde{f}_\low^\CI(\bark)$.  Below, in subsection \ref{S:lower_bound}, we specify $\alpha_\low$ and $\Gamma_\low$.   $e_\low$ is defined by
\begin{gather} \label{E:def_k_low}
e_\low = \inf \{\bark : P(\tilde{f}_\low^\CI(\bark) \ge \hat{f}) \ge .05)\}.
\end{gather}
Importantly, $\alpha_\low$ and $\Gamma_\low$ guarantee the stochastic bound,
\begin{equation} \label{E:f_upper_bound}
\tilde{f}_\p^\CI(\bark) \le \tilde{f}_\low^\CI(\bark),
\end{equation}
for any mapping $\p$ from $\mathbb{R}$ to $\P(\mu, t_A, P_A, \bark)$.  
	
	Now, still for fixed $\mu, t_A, P_A$, let $r_\actual(t), k_\actual(t), X_\actual(t; t_0)$ represent the values of $r(t), k(t), X(t;t_0)$ from which the data is sampled and let $\bark_\actual$ be the average of $k_\actual(t)$ on $[t_A, t_F]$.  In the context of inference, we do not know these actual parameters.  Suppose that $\hat{f}$ has a p-value of greater than $.05$ given $\tilde{f}_\actual^\CI$, i.e. the $\tilde{f}^\CI$ generated by setting $\Gamma$ and $\alpha$ according to $r_\actual(t)$, $k_\actual(t)$ and $X_\actual(t; t_0)$.  For the model based CI, assuming the actual parameters agree with the model parameters defined by $\p_\model(\bark_\actual)$ gives
\begin{equation}
\bark_\actual \in [e_{\model,1}, e_{\model,2}].
\end{equation}
	
	For the lower-bound CI, by (\ref{E:f_upper_bound})
\begin{align} \label{E:key_P_bound}
P(\tilde{f}_\low^\CI(\bark_\actual) \ge \hat{f})
\ge P(\tilde{f}_\actual^\CI(\bark_\actual) \ge \hat{f}) \ge .05.
\end{align}
Combining (\ref{E:def_k_low}) and (\ref{E:key_P_bound}) gives
\begin{equation}  \label{E:the_inequality}
e_\low \le \bark_\actual,
\end{equation}
showing that our lower-bound CI contains $k_\actual$.  Importantly, while the model-based CI assumes that the actual parameters are given by $\p_\model(\bark_\actual)$, the lower bound CI only assumes the actual parameters are contained in $\P(\mu, t_A, P_A, \bark_\actual)$.  
	
\subsubsection{Details for Model-Based CI} \label{S:model_based}
	
	To define the model-based CI, we need to specify $r_\model(t)$ for $t < t_A$, $k_\model(t)$, and $X_\model(t;t_0)$ for $t < t_A$.
		
	The expansion of viral load in both SIV and HIV infection has been observed to be roughly exponential \cite{Ribeiro_2010_J_Virol}. If we assume that the expansion in the number of infected cells is also roughly exponential, then $r(t)$ is constant.  Letting $r_0$ be the constant growth rate, $r_0$ must satisfy:
\begin{equation} \label{E:r_0}
r_0 =  \frac{1}{t_A} \log(P_A).
\end{equation}
For the model-based CI, we set
\begin{equation} \label{E:r_model}
r_\model(t) = r_0.
\end{equation}   

	 The assumption that $r(t)$ is constant should be viewed with caution for several reasons.  During the first days of infection, viral load is below measurable levels and may not be expanding at a constant rate.  Further, typically viral load is, at best, sampled every few days, meaning that interpolated viral loads are prone to error.  The ratio between viral load and number of infected cells may change over time, for example due to immune activation, making viral load dynamics an invalid proxy for infected cell dynamics.  Finally, typically viral loads are measured in the blood and may differ from the viral loads in others compartments.

	 The dynamics of CTL kill rates in early SIV and HIV infection are not completely understood.   However, tetramer labeling and IFN-g ELISPOT assays tracking response are available, e.g \cite{Turnbull_2009_J_Immunology, Bimber_2009_J_Virol, Vanderford_2011_PLOS_Pathogens}.  For either measurement approach, a typical response is an initial spike in CTL response lasting several days followed by either a steep or gradual decline.  We can use this profile to construct a generic $k_\model(t)$.  For a given $\bark$, we set $k(t_A) = 0$, $k(t_A + 4) = 2 \bark$, and $k(t_F) = 0$; $t$ values falling inside $(t_A, t_A+7)$ and $(t_A+7, t_F)$ are linearly interpolated. Notice, $k_\model(t)$ has the required average, $\bark$.   Alternatively, $k_\model(t)$ can be specifically constructed to match tetramer data, the case for the Vanderford et. al. dataset.
	 	
	 We set $X_\model(t;t_0) = w(t)/w(t_0)$, so that $X_\model(t;t_0)$ is simply the expected value of $X(t;t_0)$. 
	
\subsubsection{Details for Lower-Bound CI} \label{S:lower_bound}

	To construct the lower bound CI, we construct upper bounds for $\Gamma$ and $\alpha$; we label these upper bounds $\Gamma_\low$ and $\alpha_\low$ since they are used to form the lower-bound CI.    If we replace $\Gamma$ and $\alpha$ by upper bounds in (\ref{E:basic_inference}), standard calculus arguments show the resulting expression is an upper bound of $f(t_F)$, leading to (\ref{E:f_upper_bound}).  
	
	From (\ref{E:m_dynamics}) we have,
\begin{equation}  \label{E:initial_00}
\Gamma
= \int_0^{t_A} dP(\mu w(s) ds) \frac{1}{w(t_A)} X(t_A;s).
\end{equation}
Our first step in constructing $\Gamma_\low$ is to remove dependence on $X(t_A; s)$. In the appendix, section \ref{S:A_upper_bound}, we prove the following stochastic bound which applies to any random variable with values in $\{0, 1, 2,\dots,\}$,
\begin{equation}  \label{E:one_over_U}
X(t_A;s) \le \frac{E[X(t_A;s)]}{U},
\end{equation}
where $U$ is a uniform random variable on $[0,1]$.
Plugging this bound into (\ref{E:initial_00}) and recalling $E[X(t_A;s)] = w(t_A)/w(s)$ for $t < t_A$ gives
\begin{equation}  \label{E:initial}
\Gamma
\le \int_0^{t_A} dP(\mu w(s) ds) \frac{1}{w(s)} \left(\frac{1}{U}\right).
\end{equation}
Note that every time the Poisson process jumps, an independent sample of $U$ must be drawn.  

	The integral in (\ref{E:initial}) is easier to work with when the Poisson process has a constant rate.  This can be achieved by defining $\W(t) = \int_0^t ds w(s)$ and applying the substitution $q = \W(s)$.  Further substitutions and analysis, detailed in appendix section \ref{S:deriving_G}, lead to 
\begin{equation}  \label{E:mw_initial_bound}
\Gamma = \int_0^{\W(t_A)} P(\mu dq) 
\left(\frac{1}{1 + r_0 q} \right)
G(s(q)) \left(\frac{1}{U}\right),
\end{equation}
where $s(q) = \W^{-1}(q)$, $r_0$ is defined in (\ref{E:r_0}), and $G(t)$ can be characterized through the jump times of a Poisson process as follows.   Starting at time $t$, run a Poisson process with rate $r(t)$ backwards in time from $t$ to $0$.   Let $T$ represent the duration of time before the first jump, but if there are no jumps before time $0$, set $T = t$. For example, if the jump occurs at time $t-c$ then we have run the process $c$ time units before the jump occurred, making $T = c$.  This example assumes that $c < t$.   Then,
\begin{equation}  \label{E:G_1}
G(t) = \frac{1}{w(t)} + r_0 E[T].
\end{equation}

	From (\ref{E:G_1}) we have the immediate bound, $G(t) \le 1 + r_0 t$, which is optimal because setting $r(t) = 0$ on $[\epsilon, t_A]$ gives $G(t) \approx 1 + r_0 t$ for any $t \in [\epsilon, t_A]$.  However, such an $r(t)$ is not realistic for SIV.   To deal with such cases, and lead to an improved bound on $G(t)$, we introduce $r_\min(t)$ defined by linearly interpolating between the four points $r_\min(0) = 0$, $r_\min(t_A/3) = 3r_0/4$, $r_\min(2t_A /3) = 3r_0/4$ and $r(t_A) = 0$.  The form of $r_\min(t)$ is somewhat arbitrary.  Intuitively, we suppose that growth rates may be low at initial infection or near peak viral load, so at the endpoints we choose $r_\min(t) = 0$ and in between we assume growth rates are at least $3/4$th of the average growth rate, $r_0$.   As we lower this fraction, say from $3/4$ to $1/2$, the bound of $G(t)$ becomes poorer.
	
	We require $r(t) \ge r_\min(t)$ for $r(t)$ to be in $\P(\mu, t_A, P_A, \bark)$.  Given this constraint, we can use $r_\min(t)$ as a worst case and derive the bound,
\begin{equation}  \label{E:mw_bound_2}
\Gamma \le  \int_0^{\W(t_A)} P(\mu dq) 
\left(\frac{1}{1 + r_0 q} \right)
G_\max(s(q)) \left(\frac{1}{U}\right),
\end{equation}
where $G_\max(t)$ is constructed using $r_\min(t)$ to define $T$ and the $w(t)$ in (\ref{E:G_1}) is replaced by $w_\min(t)$, where $w_\min(t) = \int_0^t ds r_\min(s)$.

	The right hand side of (\ref{E:mw_bound_2}) is almost independent of $r(t)$, except for the boundary point of integration, $\W(t_A)$.   Recall that $w(t_A) = P_A$ and $\W(t_A)$ is the integral of $w(t)$ from time $0$ to $t_A$.   Intuitively, the faster the wild-type population expands, the more mutations will occur; $\W(t_A)$ quantifies this intuition.   Generally, the best bound of $\W(t_A)$ is $\W(t_A) \le P_A t_A$ which is achieved if the viral population instantly reaches size $P_A$ and then stays fixed.  Clearly this is not biologically realistic, so we introduce $w_\max(t)$ defined by the growth rate $r_\max(t)$,
\begin{equation}
r_\max(t) = 2 \bark (t_A - t),
\end{equation}
and require that $r(t) \in \P(\mu, t_A, P_A, \bark)$ satisfy $w(t) \le w_\max(t)$ for $t \le t_A$ (recall that $w(t)$ for $t < t_A$ is parameterized by $r(t)$ through (\ref{E:w_dynamics})).  Letting $\W_\max(t_A)$ be the integral of $w_\max(t)$, we can finally define,
\begin{equation}
\Gamma_\low =  \int_0^{\W_\max(t_A)} P(\mu dq) 
\left(\frac{1}{1 + r_0 q} \right)
G_\max(s(q)) \left(\frac{1}{U}\right),
\end{equation}
which satisfies the stochastic bound, $\Gamma \le \Gamma_\low$.

	To form an an upper bound for $\alpha$, we construct a $k_\low(t)$ that maximizes $\alpha$ for fixed $\bark$.  With solely the restriction $k(t) \ge 0$, no such maximum exists. For example, letting
\begin{equation}
k(t) = \bigg\{
\begin{array}{cc}
0 & \text{for } t < t_F - \epsilon \\
\frac{2 \bark}{\epsilon} (t_F - t) & \text{for } t \ge t_F - \epsilon, 
\end{array}
\end{equation}
then $\alpha \to \infty$ as $\epsilon \to 0$ with $\bark$ remaining fixed.   

	To avoid such pathological cases, we assume $k(t)$ is formed through two piece-wise linear interpolations.  More precisely, we allow $k(t)$ to take any non-negative values at times $t_A, t^*, t_F$ where $t^*$ is any time in $[t_A, t_F]$.  For other times $t \in [t_A, t_F]$, $k(t)$ is formed by linearly interpolating based on $k(t_A), k(t^*)$, and $k(t_F)$.    Since $t_F - t_A$ is less than $2$ weeks for the datasets we consider, this parametrization of $k(t)$ allows for the range of kill rate profiles seen in experimental datasets.  We could define $k(t)$ by linearly interpolating more than three points; but as the number of such points rises, our bound of $\alpha$ will become poorer.

	With $\bark$ fixed, set
\begin{equation}
k_\low(t) = 2 \bark \left(\frac{t - t_A}{t_F - t_A}\right)
\end{equation}
Let $\alpha_\low(\bark)$ be the value of $\alpha$ based on  $k_\low(t)$, and let $\alpha(\bark)$ correspond to any $k(t)$ satisfying our two piece linear interpolation assumption and having average $\bark$.  Then standard optimization arguments show,
\begin{equation}
\alpha(\bark) \le \alpha_\low(\bark).
\end{equation}
Intuitively, $k_\low(t)$ is a CTL response that expands slowly and such response allows more wild types to mutate during $[t_A, t_F]$ leading to a higher fraction of mutants at $t_F$.

\begin{figure} [h]
\begin{center} 
\includegraphics[width=1\textwidth]{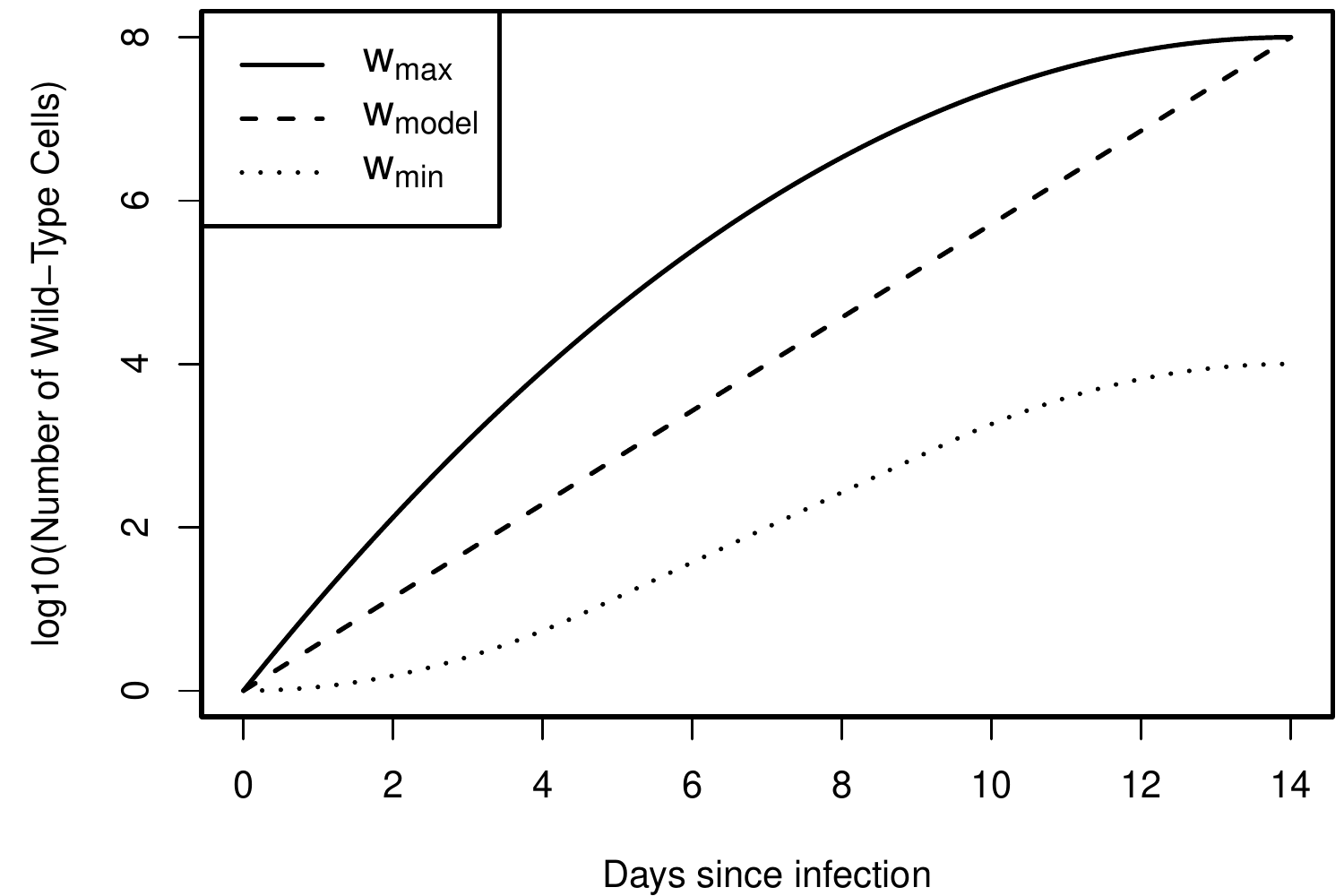}
\caption{Constraints on $r(t)$ and $w(t)$} 
\label{F:figure_w}
\end{center}
\end{figure}

	In the preceding paragraphs we've specified two additional constraints on $\P(\mu, t_A, P_A, \bark)$ beyond those already mentioned:
\begin{itemize}
\item $r(t) \in \P(\mu, t_A, P_A, \bark)$ if for all $t \le t_A$: $w(t) \le w_\max(t)$ and $r(t) \ge r_\min(t)$,
\item $k(t) \in \P(\mu, t_A, P_A, \bark)$ if $k(t)$ can be constructed through the linear interpolation of the three points $k(t_A), k(t^*)$, and $k(t_F)$ with $t^* \in [t_A, t_F]$.  
\end{itemize}
Figure \ref{F:figure_w}, which is constructed assuming $t_A = 14$ and $P_A = 10^8$, gives an intuition for the constraints on $r(t)$.  Roughly, $r(t) \in \P(\mu, t_A, P_A, \bark)$ if the $w(t)$ it parametrizes falls between $w_\max(t)$ and $w_\min(t)$.  This is not quite right, because actually we require $r(t) \ge r_\min(t)$ rather than $w(t) \ge w_\min(t)$, but the figure gives an intuition for the range of acceptable $r(t)$.    Seven days into infection, $w_\max$ and $w_\min$ have log10 values of $6$ and $2$, showing that there is substantial variation in acceptable $r(t)$.   Notice that $r_\model(t)$ is roughly a middle ground.    

\subsubsection{Dataset Parameters}  \label{S:parameters}

	For both Bimber et. al and Vanderford et. al. datasets, we set $n$ in (\ref{E:tilde_f}) to $100$.   If each sequence samples from the viral population independently, $n$ represents the number of sequences sampled.   Since both datasets are generated through deep sequencing, their sequence counts are in the $1000$s; however, pyrosequencing is subject to biasing, so that the effective sampling count is likely much lower.   We set $n=100$, reflecting a conservative assumption of frequency estimation with roughly $10\%$ accuracy.   
	
	$r_\model(t)$ and $X_\model(t;t_0)$ have the generic form specified in section \ref{S:model_based} for both datasets.  For Bimber et. al., $k_\model(t)$ also takes the generic form mentioned in that section: the linear interpolation of $k(t_A) = 0$, $k(t_A + 4) = 2 \bark$, and $k(t_F) = 0$.   For Vanderford et. al., the $k_\model(t)$ used in the LN compartment is the linear interpolation of $k(t_A) = 0$ and $k(t_F) = 2\bark$, while the $k_\model$ in the RB compartment takes the generic form used in Bimber et. al.
	
	Values for $\mu, t_A, P_A$ and $t_F$ are given in the Results section.

\subsubsection{Fitness Costs} \label{S:fitness}
	
	To model fitness costs, we change the expected value of $X(t;t_0)$ from (\ref{E:X_E}) to
\begin{align} \label{E:X_E_fit}
E[X(t;t_0)] & = \exp[\int_{t_0}^t ds (r(s) - \c)]
\\ \notag
	& = \exp[-\c(t-t_0)]
\left(\frac{w(t)}{w(t_0)}\right) \exp[\int_{t_0}^t ds k(s)],
\end{align}
where $\c$ is the fitness cost associated with mutation.  Changing $X(t;t_0)$ alters the dynamics of $m(t)$, although (\ref{E:m_dynamics}) still holds.  

	For a moment, assume that we are given $\c$ and would like to estimate $\bark$.  We can apply (\ref{E:basic_inference}) as before, but with $\Gamma, z$ and $\alpha$ replaced by corresponding versions defined under the fitness cost model: $\Gamma_\fit$, $z_\fit$, and $\alpha_\fit$.  Assuming that mutants are less fit gives the stochastic bound $\Gamma_\fit \le \Gamma$, while $z_\fit$ and $\alpha_\fit$ equal $z, \alpha$ under the transformation $k(t) \to k(t) - \c$.  More precisely,
\begin{equation} 
z_\fit = \exp[-(\bark - \c)(t_F - t_A)],
\end{equation}
and
\begin{equation} 
\alpha_\fit = \mu \int_{t_A}^{t_F} ds \exp[\int_s^{t_F} ds' (k(s') - \c)].
\end{equation}

	Let $[e_{\model,1}^\fit, e_{\model,2}^\fit]$ and $[e_\low^\fit,\infty)$ be the model-based and lower-bound CIs constructed under the fitness model with $\c$ given.   Using the same arguments and notation stated in the paragraph below (\ref{E:f_upper_bound}) gives
\begin{align}  \label{E:fit_e}
\bark_\actual \in [e_{\model,1}^\fit, e_{\model,2}^\fit]
\end{align}
if the model parameters and actual parameters are equal, and
\begin{align}  \label{E:fit_e_low}
\bark_\actual \in [e_\low^\fit, \infty)
\end{align}
if the actual parameters are in $\P(\mu, t_A, P_A)$.
	
	Now suppose that we construct our CIs under the model assuming equal fitness, but that $\hat{f}$ is drawn from $f(t_F)$ under the fitness cost model.  (Recall that the model-based and lower-bound CIs under the equal fitness model are $[e_{\model,1}, e_{\model,2}]$ and $[e_\low,\infty)$, respectively.)  The stochastic bound, $\Gamma_\fit \le \Gamma$ gives
\begin{equation}  \label{E:fit_basic_bound}
\frac{1}{1 + \frac{z_\fit}{\Gamma_\fit + z_\fit \alpha_\fit}} \le \frac{1}{1 + \frac{z}{\Gamma + z \alpha}},
\end{equation}
where $z_\fit$ and $\alpha_\fit$ are formed with kill rate $k(t)$ while $z$ and $\alpha$ are formed with kill rate $k(t) - \c$.  The expression to the left and right sides of the $\le$ in (\ref{E:fit_basic_bound}) are used to form CIs under the fitness cost and equal fitness models, respectively.   Exploiting (\ref{E:fit_basic_bound}) as we did (\ref{E:f_upper_bound}), gives :
$e_{\model,1} + \c \le e_{\model,1}^\fit$,
$e_{\model,2} + \c \le e_{\model,2}^\fit$ and
$e_\low + \c \le e_\low^\fit$.

	By combining (\ref{E:fit_e}) and (\ref{E:fit_e_low}) with the relations directly above, we find
\begin{equation}  \label{E:fit_model_final}
\bark_\actual - \c \ge e_{\model,1},
\end{equation}
assuming the actual parameters and model parameters are equal, and
\begin{equation}  \label{E:fit_low_final}
\bark_\actual - c \in [e_\low, \infty)
\end{equation}
assuming the actual parameters are in $\P(\mu, t_A, P_A)$.

	When fitness costs are included, the escape rate is given by the difference between the average kill rate and the mutation fitness cost \cite{Asquith_PLOS_Biology_2006}.  From (\ref{E:fit_low_final}), the lower-bound CI always includes the escape rate, even though the CI is formed under a model assuming equal fitness.  From (\ref{E:fit_model_final}), the model-based CI may not contain the escape rate, but the escape rate is always above the left endpoint of the CI.

\subsection{Numerical Experiments}

\begin{figure} [h]
\begin{center} 
\includegraphics[width=1\textwidth]{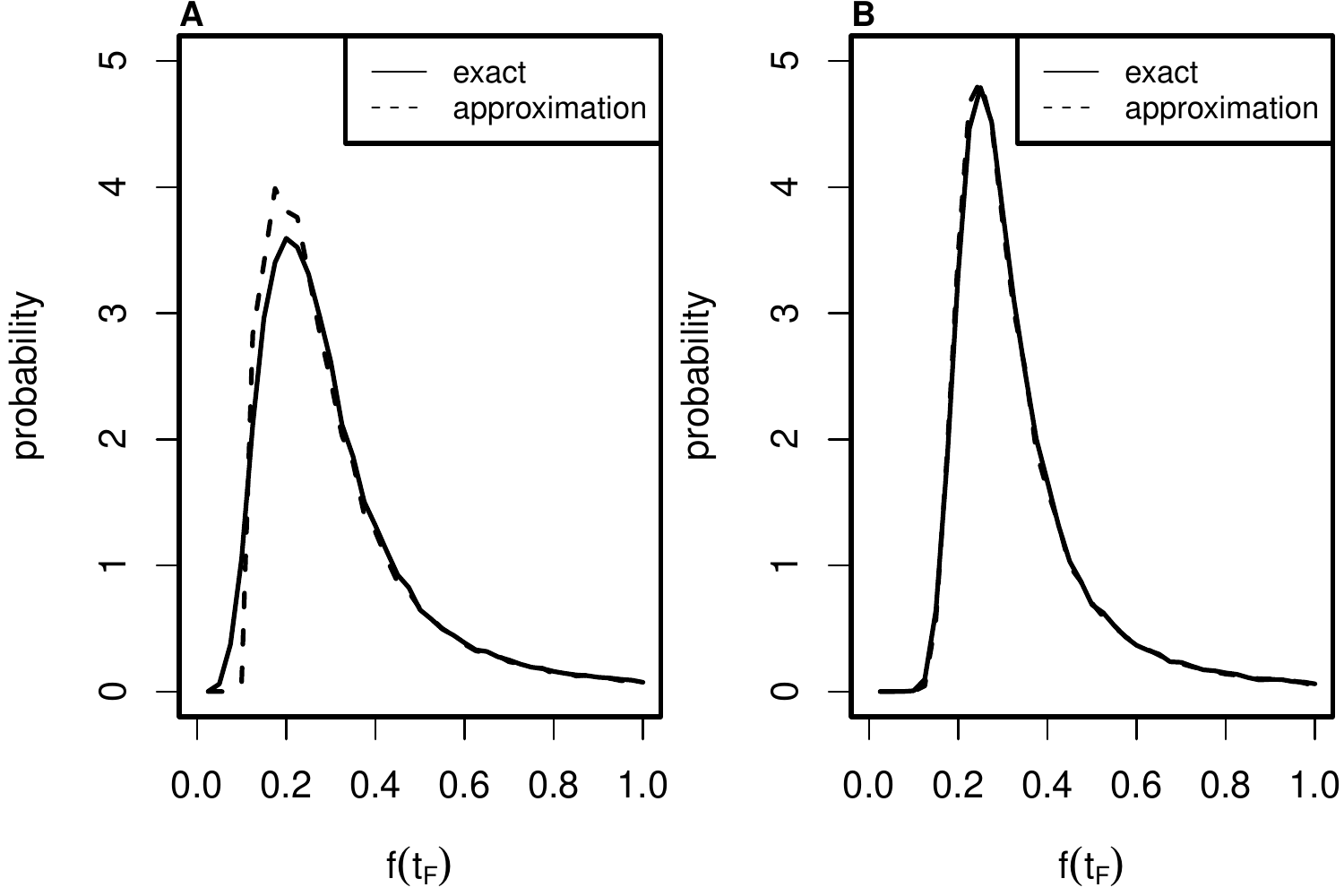}
\caption{Accuracy of (\ref{E:basic_inference}). The subfigures show cdfs for $f(t_F)$ based on exact simulations and the approximation (\ref{E:basic_inference}).   For both subfigures, $t_A = 9$ and $t_F = 14$, but in (A) $P_A = 7200$ and (B) $P_A = 32000$.}
\label{F:alpha_Vand}
\end{center}
\end{figure}

	Through simulation, we investigated the accuracy of estimating $f(t_F)$, the mutant frequency at sampling time, through (\ref{E:basic_inference}).  Figure \ref{F:alpha_Vand} considers a setting similar to the lymph node compartment of the Vanderford et. al. dataset, with relatively early CTL response and sampling times: $t_A = 9$ and $t_F = 14$.   For subfigures $A$ and $B$ we imagined peak viral loads at day 14 with $10^6$ and $10^7$ infected cells, and based on these values we interpolated the number of infected cells at $t_A$ by assuming a constant growth rate; this gave $P_A$ values of roughly $7200$ and $32000$, respectively.  The subfigures show empirical cdfs constructed from $10^5$ simulations, the exact cdf is generated by simulating the model while the approximate cdf uses (\ref{E:basic_inference}).   The growth and kill rates were those used to construct the model-based CI for the Vanderford et. al. dataset, with $\bark$ set at $1$.   
	
	The approximating cdfs in both subfigures are relatively accurate, although the larger $P_A$ in subfigure B improves the accuracy.   With $P_A = 32000$ the mutation rate at time $t_A$, given by $\mu P_A$, is roughly $9$.  Generally, across different numerical experiments (data not shown), we found that a mutation rate at time $t_A$ greater than $5$ led to an accurate approximation of $f(t_F)$ by (\ref{E:basic_inference}). Mutation rates at $t_A$ are on the order of $10^4$ for our Bimber et. al. parametrization.    
	
	To assess the sensitivity of our CIs  to the parameters $\mu, t_A, P_A$; the growth and kill rates, $r(t), k(t)$; and the offspring distribution $X(t_A;t)$; we used the following testing approach.
\begin{enumerate}
\item We generated simulated mutation frequencies at day $t_F$ using our model and a choice for the parameters.   We refer to these parameters as the \textit{simulation parameters}.
\item Given the simulated mutation frequencies at day $t_F$, we chose parameter values and built the model-based and lower-bound CIs.   We refer to these parameters as the \textit{inference parameters}.
\end{enumerate}

\begin{figure} [h]
\begin{center} 
\includegraphics[width=1.2\textwidth]{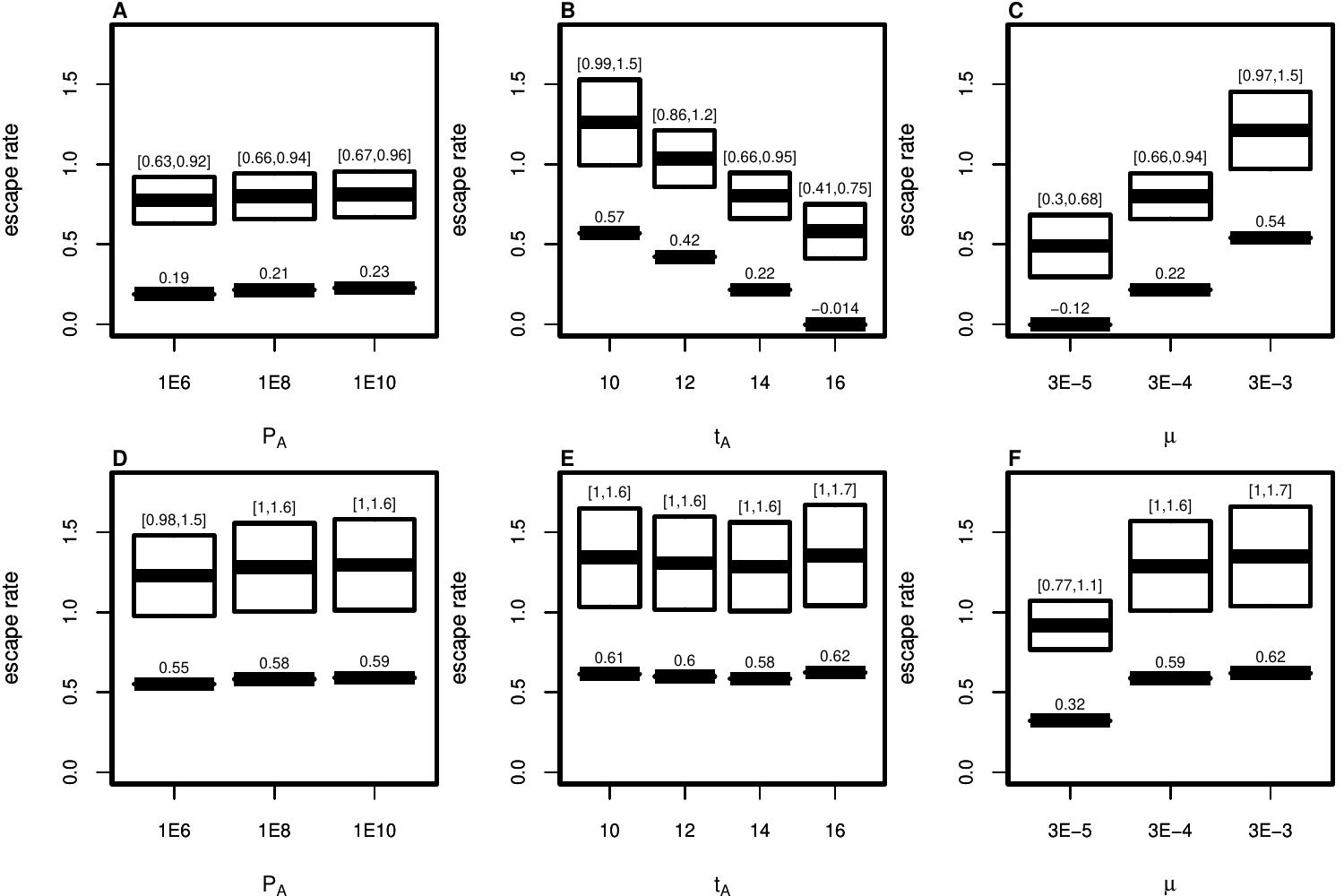}
\caption{Results of numerical experiments.  In each subfigure, one parameter was varied.   Subfigures A,B,C show experiments in which the CI is inferred with the same $r(t), k(t), X(t;t_0)$ values as used for generating simulated mutation frequencies, in subfigures D,E,F the $r(t), k(t), X(t;t_0)$ differ between inference and simulation.  See text for further details.} 
\label{F:numerical_Bimber}
\end{center}
\end{figure}

Figure \ref{F:numerical_Bimber} shows results with $t_F = 21$.  Each tic in each subfigure was generated by repeating the two step, testing approach $1000$ times and averaging.   

	For all subfigures, our inference parameters were fixed at $\mu = 3*10^{-4}$, $t_A = 14$, $P_A = 10^8$ and $r(t) = r_\model(t)$, $k(t) = k_\model(t)$, $X(t_A;t) = w_\model(t_A)/w_\model(t)$.  Intuitively, given data we always guessed the same parameters; but, importantly, to generate the lower-bound CI only the values of $\mu, t_A, P_A$ are needed.   For the simulation parameters, in all subfigures we fixed $\mu, t_A, P_A$ to the same values as the inference parameters, except that in each subfigure the x-axis shows how we varied a single one of these parameters.  For example, in subfigure A we simulated the data using $\mu = 3*10^{-4}$, $t_A = 14$, and with $P_A$ taking on the values $10^6, 10^8, 10^{10}$ over three separate numerical experiments.    For figures A,B,C we set $r(t), k(t), X(t_A;t)$ to the same values as the inference parameters, while for figures D,E,F the values were different (see the Appendix section \ref{S:A_nm} for the specific values we chose); so what distinguishes subfigure A from D is whether or not the inference  and simulation parameter values for $k(t), r(t), X(t_A;t)$ are equal and similarly for subfigures B,E and C,F.  

	In all experiments we set $\bark = .8$ (the $k(t)$ chosen as both simulation and inference parameters depend on $\bark$).  In subfigures A,B,C the tics $P_A=10^8$, $t_A=14$ and $\mu = 3*10^{-4}$, respectively, are the same experiment in which simulation and inference parameters are equal.   For this case, the model based-CI contains the true $\bark$ value.   Subfigure A shows that varying $P_A$ has little effect on our model-based CI, while the other subfigures show that varying any other parameter leads to an erroneous CI.  This, however, is not the case for the lower-based CI which contains the true $\bark$ over all parameter choices.
	
	Figure \ref{F:numerical_Bimber} shows averaged CI.  Of the $1000$ experiments run to produce each tic, the model-based CI contained $\bark$ approximately $92\%$ of the time when the simulation and inference parameters were identical or only $P_A$ was varied as in subfigure A.   For all other tics, the model-based CI contains $\bark$ less than $25\%$ of the time.  In contrast, over all tics and all $1000$ experiments for each tic, the lower-bound CI always contained $\bark$. 
	
	We also ran experiments in which no simulated and inference parameters agreed.   The inference parameters were unchanged from the previous experiments, but the simulated parameters were set at the values for which the lower bound was highest in Figure \ref{F:numerical_Bimber} ($\mu = 3*10^{-3}$, $P_A = 10^{10}$, $t_A = 16$, and  $r(t), k(t), X(t;t_0)$ that of subfigures D,E,F in Figure \ref{F:numerical_Bimber}).   Over $1000$ experiments, no model-based CI contained $\bark$, while the lower-bound CI always contained $\bark$.  
	
	As can be seen from Figure \ref{F:numerical_Bimber}, the lower-bound CI significantly underestimates $\bark$ (recall we set $\bark=.8$ for the numerical experiments).  Some of this underestimate comes from model uncertainty.   In forming the lower-bound CI, we constructed $\alpha_\low$ and $\Gamma_\low$ as upper bounds for $\alpha$ and $\Gamma$.   $\alpha_\low$ is an optimal bound because $k_\low(t) \in \P(\mu, t_A, P_A, \bark)$, so when $k(t) = k_\low(t)$ we have $\alpha = \alpha_\low$.  In situations when $\alpha_\low > \alpha$, the lower-bound CI will underestimate $\bark$ as a consequence of our uncertainty regarding which $k(t) \in \P(\mu, t_A, P_A, \bark)$ generated the data.   But the situation is different for $\Gamma_\low$ because no $r(t)$ can simultaneously satisfy $r(t) = r_\max(t)$ and $r(t) = r_\min(t)$ and no $X(t;t_0)$ can satisfy $X(t;t_0) = E[X(t;t_0)]/U$.  When we simulate data using $p_\model$ (meaning we use $r_\model(t)$, $k_\model(t)$, $X(t;t_0) = E[X(t;t_0)]$) and take $k_\model(t) = k_\low(t)$, we find that approximately $90\%$ of the distance between the lower bound and the correct confidence interval comes from bounding $X(t;t_0)$ by $E[X(t;t_0)]/U$ while $10\%$ comes from the use of $r_\min$ and $r_\max$.  This is the case, for example, for the second tic of Figure \ref{F:numerical_Bimber}A, in which the lower bound is $.21$ and the correct CI is $[.66, .94]$.

\section{Discussion}

	Inferring the rate of the first CTL escape involves a tradeoff.  One one hand, the existing two-point method is largely model independent but can only be applied using two sampled timepoints, meaning that the early part of the escape is often missed and inference implicitly focuses on later parts of the escape.   On the other hand, if a more parameterized method is used, the early part of the escape can be considered but inference results depend on model structure and the parameter values chosen.
	
	In this work, we have developed escape rate inference methods applicable to single timepoint datasets with an effort to minimize model dependence and the number of parameters.   To do this, we construct lower-bound confidence intervals which account for model uncertainty, although three parameters must still be specified.  
	
	As our results show, in some cases these lower-bounds are not informative, but in many cases the lower-bounds allow us to understand CTL escape in a more dynamic fashion.  In the Bimber et. al. dataset, lower-bound CIs combined with the two-point method reveal faster rates for the first CTL escape during days 14-21 than days 21-28. The Vanderford et. al. data has a similar pattern, with faster rates of escape in the lymph nodes during days 9-14 than during days 14-28.  Further, we were able to show that escape in the rectal mucosa is at least as fast, and likely faster, than lymph node escape during days 14-28.   
	
	The lower-bound CI accounts for uncertainty in the population dynamics of SIV infected cells and CTL response, however our methods still depend on the parameters $\mu, t_A$ and $P_A$.   We made the choice $\mu = 3*10^{-4}$, as mentioned in \cite{Mandl_J_Virol_2007} the mutation rate of TAT-SL8 is possibly lower but unlikely to be higher.   A lower value for $\mu$ would raise our escape rate CIs, leaving our conclusions unchanged.   We chose $t_A$, the time CTL response begins, based on tetramer data for the percentage of CD8+ lymphocytes recognizing the epitope.  To the extent such data reflects underlying CTL kill rates, our choice of $t_A$ is justified.   Finally, numerical experiments show the CIs are weakly dependent on $P_A$, the number of infected cells at time $t_A$.   
	
\begin{figure} [h]
\begin{center} 
\includegraphics[width=1.2\textwidth]{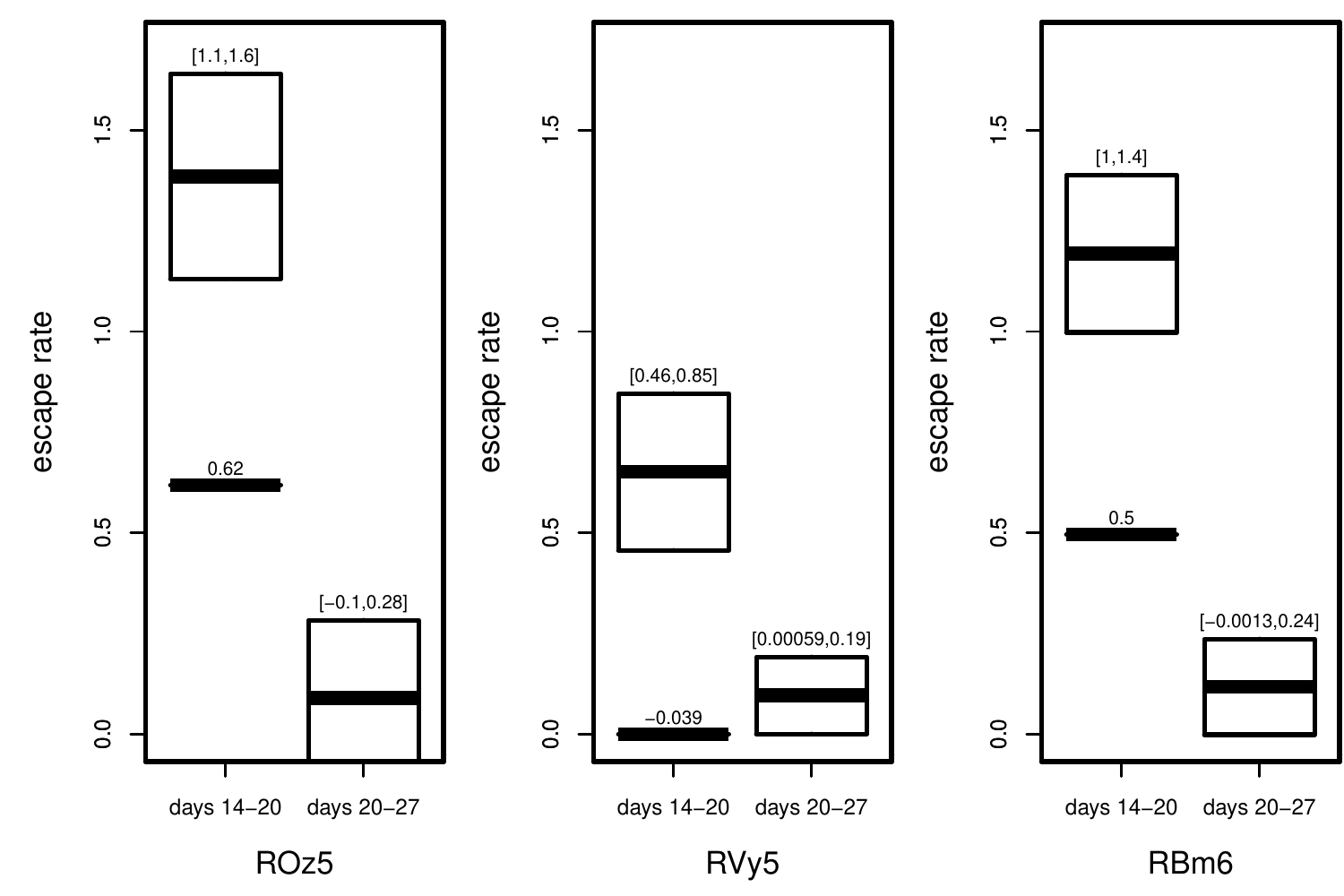}
\caption{Escape Rates for Mandl et. al.} 
\label{F:Mandl}
\end{center}
\end{figure}

	Our estimates are in line with previous work.  Mandl et. al. \cite{Mandl_J_Virol_2007} considered escape at TAT-SL8 in rhesus macaques using a standard-model based inference method.  By fitting the standard-model using six timepoints, they estimated escape rates in the range of $.3$ to $.9$.   We applied our methods to their dataset, see Figure \ref{F:Mandl}, finding lower bound CIs consistent with the estimates of Mandl et. al. Further, in two of the three animals, lower-bound CI for the escape rate during days 14-20 were greater than the two-point method CI for the days 20-27 escape rate.   The exception, animal RVy5, had a weaker TAT-SL8+, CTL response prior to day 20 than the other two animals.   As in Bimber et. al., an association seems to exist between the strength of CTL response and early escape rates.  However, these associations are not statistically significant and more work is required.  
	
	In this work, we have not distinguished between different types of epitope mutations, although CTL escape typically involves multiple mutation variants \cite{Boutwell_2010_J_Infec_Dis}.  As CTL binding affinity may differ between mutation variants, we are really inferring an average escape rate over all mutations at the epitope.   For TAT-SL8, most mutations arising in escape have low binding affinity, so our estimates may apply without much modification \cite{Allen_2000_Nature, OConner_2002_Nature}. 
	
	As discussed in the Methods section, the lower-bound CI contains the escape rate even when mutations incur a fitness cost.   Reversion studies suggest TAT-SL8 mutations may not be associated with a fitness cost, but mutations at other epitopes are often less fit \cite{Goulder_Nature_Reviews_2004, Friedrich_2004_Nature_Med}.    
	
	The lower-bound CI as currently constructed is overly conservative.  Often, as demonstrated by numerical experiments in the Methods section, the lower-bound CI significantly underestimates the escape rate even when model uncertainty is accounted for.  Improving the lower-bound CI requires better quantitative understanding of acute infection.   For example, the number of offspring infected cells descendant from a single HIV or SIV infected cell is not well understood.   Estimates of $R_0$ provide averages, but better lower-bounds require knowledge of the distribution.   Since we lack such estimates we are forced to use a crude bound, i.e. $X(t;t_0) \le E[X(t;t_0)]/U$, leading to overly conservative lower-bound CIs.  Future work leading to improved lower-bound CIs would greatly expand the range of datasets we could consider as well as sharpen our current estimates.

\section*{Acknowledgements}
I would like to thank Guido Silvestri and Thomas Vanderford for providing the Vanderford et. al. dataset analyzed in this paper.  I would like to thank Shelby O'Connor for providing several datasets that helped me understand early SIV infection, and also for answering several questions relating to the Bimber et. al. dataset and SIV infection in general.   This work was supported by NSF grant DMS-1225601.

\appendix
\section{Appendix}
\renewcommand{\theequation}{A.\arabic{equation}}
\setcounter{equation}{0}

\subsection{Derivation of (\ref{E:mw_initial_bound}) and (\ref{E:G_1})}  \label{S:deriving_G}

To see (\ref{E:mw_initial_bound}) and (\ref{E:G_1}), first apply the substitution $q = \W(s)$ to the integral in (\ref{E:initial}), giving
\begin{equation}  \label{E:G_der_0}
\Gamma
= \int_0^{\W(t_A)} P(\mu dq) \frac{1}{w(s(q))} \left(\frac{1}{U}\right).
\end{equation}
When $r(t) = r_0$, a simple computation shows $w(s(q)) = 1 + r_0 q$.   To understand deviations from constant $r(t)$ we rewrite (\ref{E:G_der_0}) using the expression $1+r_0q$,
\begin{equation}  \label{E:G_der_1}
\frac{m(t_A)}{w(t_A)}
= \int_0^{\W(t_A)} P(\mu dq) \frac{1}{1 + r_0 q} 
\left[G(s(q))\right] \left(\frac{1}{U}\right),
\end{equation}
where $G(s(q)) = (1 + r_0q)/w(s(q))$ but can be profitably expressed as 
\begin{equation}  \label{E:G_der_2}
G(t) = \frac{1}{w(t)} + r_0 \int_0^t ds 
\exp[-\int_{s}^t ds' r(s')].
\end{equation}
(\ref{E:G_der_2}) can be seen by noting
\begin{equation}
q(t)/w(t) = \W(t)/w(t) = \int_0^t ds w(s)/w(t),
\end{equation}
and then replacing $w(s)$ by $\exp[\int_0^s ds' r(s')]$ and similarly for $w(t)$.  

	Applying integration by parts to the integral in (\ref{E:G_der_2}) we have,
\begin{align}
G(t) & = \frac{1}{w(t)} + r_0 
\left[ -\int_0^t ds (s) r(s) \exp[-\int_{s}^t ds' r(s')]
+ s \exp[-\int_{s}^t ds' r(s')]\bigg|_0^t] \right]
\\ \notag
	& = \frac{1}{w(t)} + r_0
	\left[ -\int_0^t ds (t - (t-s)) r(s) \exp[-\int_{s}^t ds' r(s')] + t \right]
\\ \notag
& = \frac{1}{w(t)} + r_0
	\left[-t(1 - \frac{1}{w(t)}) + \int_0^t ds (t-s) r(s) \exp[-\int_{s}^t ds' r(s')] + t \right]
\\ \notag
& = \frac{1}{w(t)} + \frac{1}{w(t)} r_0 t
+ \int_0^t ds (t-s) r(s) \exp[-\int_{s}^t ds' r(s')].
\end{align}
In the last expression, $r(s) \exp[-\int_s^t ds' r(s')]$ can be seen as the non-normalized pdf for the jump time of a Poisson process run at rate $r(t)$ from $t$ backwards in time.   Now the integral of the pdf, $\int_0^t ds r(s) \exp[-\int_{s}^t ds' r(s')]]$, evaluates to $1 - 1/w(t)$.   We can then write,
\begin{equation}
G(t) = \frac{1}{w(t)} + r_0 \left(\frac{t}{w(t)} 
+ (1 - \frac{1}{w(t)})
\int_0^t ds (t-s) r(s) \frac{\exp[-\int_{s}^t ds' r(s')]}{1 - \frac{1}{w(t)}} \right)
\end{equation}
which is exactly the statement (\ref{E:G_1}).

\subsection{Proof of (\ref{E:one_over_U})} \label{S:A_upper_bound}

For the sake of generality we consider a random variable $Y$ with the restriction that $Y \in \{0,1,2,\dots\}$.   Throughout, we use $y$ to represent a scalar and $Y$ to be the random variable.  First,  letting $F(y)$ be the cdf of $Y$, we show that for $y \ge 1$,
\begin{equation}  \label{E:a_first_bd}
y \le \frac{E[Y]}{1 - F(y-1)}.
\end{equation}
Indeed, this is just a Chebyshev bound. 
\begin{equation}
y(1 - F(y-1)) = y P(Y \ge y) \le E[Y].
\end{equation}

Next we define a r.v. $H$ by,
\begin{equation}
H = \bigg\{
\begin{array}{cc}
1 - F(Y-1) & \text{if } Y \ne 0 \\
1 & \textbf{otherwise}.
\end{array}
\end{equation}  
Then using (\ref{E:a_first_bd}) and the definition of $H$ gives,
\begin{equation}
Y \le \frac{E[Y]}{H}
\end{equation}
We're done if we can show that $H \ge U$ where $U$ is a uniform r.v. on $[0,1]$.  To see this, set $P(Y = i) = c_i$.   Then,
\begin{align}
H = \bigg\{
\begin{array}{cc}
1 & \text{with probability } c_0 \\
1 - c_0 & \text{with probability } c_1\\
1 - c_0 - c_1 & \text{with probability } c_2\\
\vdots
\end{array}
\end{align}
To prove $H \ge U$, 
define the mapping $\Phi : [0,1] \to \{1, 1-c_0, 1 - c_0 - c_1, \dots\}$,
\begin{align}
\Phi(x) = \bigg\{
\begin{array}{cc}
1  & \text{for } x \in [1-c_0, 1] \\
1 - c_0  & \text{ for } x \in [1-c_0-c_1,1-c_0] \\
1 - c_0 - c_1 & \text{ for } x \in [1-c_0-c_1-c_2,1-c_0-c_1] \\
\vdots
\end{array}
\end{align}
Then $U \le \Phi(U) \sim H$ and we are done. 

\subsection{Details for Numerical Experiments}  \label{S:A_nm}

Here are the values for $r(t), k(t), X(t;t_0)$ used as simulation parameters to generate subfigures D,E,F in Figure \ref{F:numerical_Bimber}.
\begin{equation}
r(t) = \bigg\{
\begin{array}{cc}
.2r_0 & \text{if } t \in [0, \frac{t_A}{3}) \\
1.8 r_0 & \text{if } t \in [\frac{t_A}{3}, \frac{2t_A}{3})\\
r_0 & \text{if } t \in [\frac{2t_A}{3}, t_A]
\end{array}
\end{equation}

$k(t)$ is linearly interpolated from the following three points: $k(t_A) = 0$, $k(t_A + 4) = \tilde{k}$, $k(t_F) = \tilde{k}$ where
\begin{equation}
\tilde{k} = \bark \left(\frac{t_F - t_A}{t_F - t_A - 4}\right),
\end{equation}
and $X(t;t_0)$ is given by
\begin{equation}
X(t;t_0) = 100 \left(\frac{w(t)}{w(t_0)}\right) \textbf{Binomial}(.01; 1).
\end{equation}

\newcommand{\noopsort}[1]{} \newcommand{\printfirst}[2]{#1}
  \newcommand{\singleletter}[1]{#1} \newcommand{\switchargs}[2]{#2#1}

\end{document}